\documentclass[3p,twocolumn]{elsarticle}
 
 \usepackage{dblfloatfix}
 \usepackage[
    type={CC},
    modifier={by-nc-nd},
    version={4.0},
]{doclicense}
\usepackage{textcomp}
\usepackage{booktabs}
\usepackage{dot2texi}
\usepackage{textgreek}
\usepackage{amssymb}
\usepackage[fleqn]{amsmath}
\usepackage{amsthm}
\usepackage[table]{xcolor}
\usepackage[color]{circusPP}
\usepackage{tikz}
\usetikzlibrary{shapes,arrows}
\usetikzlibrary{shapes.geometric}
\usepackage{stmaryrd}
\usepackage{hyperref}
\usepackage{acronym}
\usepackage[capitalise]{cleveref}
\usepackage{tikz}
\usepackage[caption=false]{subfig}
\tikzset{square/.style={regular polygon,regular polygon sides=4,inner sep=0}}
\newcommand\encircle[1]{%
  \tikz[baseline=(X.base)] 
    \node (X) [draw, circle, inner sep=-0.15em,fill=black!100,text=white] {\strut #1};}

\makeatletter
\def\@endtheorem{\endtrivlist}
\makeatother

\newtheoremstyle{linked}
  {}
  {}
  {\itshape}
  {}
  {\bfseries}
  {.}
  { }
  {\href{\thislink}{\thmname{#1} \thmnumber{#2}}\thmnote{ (#3)}}

\newcommand{\isalogo}{\includegraphics[width=9pt]{isabelle_transparent}}
\newcommand{\isaicon}[1]{\href{#1}{\isalogo}}

\journal{Information Processing Letters}

\def\IINew{\hbox{\texttt{\slshape I \kern -9.5pt I}}}

\newtheorem{definition}{Definition}
\newtheorem{example}{Example}
\newtheorem{lemma}{Lemma}
\newtheorem{theorem}{Theorem}

\theoremstyle{linked}
\newtheorem{innlinkthm}[theorem]{Theorem}
\newenvironment{linkthm}[1]
 {\def\thislink{#1}\innlinkthm}
 {\endinnlinkthm}

\newtheorem{innlinklemma}[lemma]{Lemma}
\newenvironment{linklemma}[1]
 {\def\thislink{#1}\innlinklemma}
 {\endinnlinklemma}

\newtheorem{innlinkdefinition}[definition]{Definition}
\newenvironment{linkdefinition}[1]
 {\def\thislink{#1}\innlinkdefinition}
 {\endinnlinkdefinition}

\acrodef{CSP}{Communicating Sequential Processes}
\acrodef{ZRC}{Z Refinement Calculus}
\acrodef{VDM}{Vienna Development Method}
\acrodef{ASM}{Abstract State Machine}
\acrodef{FSM}{Finite State Machines}
\acrodef{CCS}{Calculus of Concurrent Systems}
\acrodef{JCSP}{Java Communicating Sequential Processes}
\acrodef{FDR}{Failures-Divergence Refinement}
\acrodef{UTP}{Unifying Theories of Programming}
\acrodef{BNF}{Backus-Naur Normal Form}

\makeatletter
\newcommand{\thenT}{\zbinop{\csym@cspthen}}
\newcommand{\extchoiceT}{\zbinop{\csym@cspextchoice}}
\newcommand{\circinterruptT}{\zbinop{\csym@circinterrupt}}
\newcommand{\circinterruptU}{\zbinop{\csym@circinterrupt_U}}
\newcommand{\rparT}{\zbinop{\zopenop{\csym@rpar}}}
\makeatother

\biboptions{sort&compress}

\begin{document}

\begin{frontmatter}

\title{Priorities in $tock$-CSP}

\author{Pedro Ribeiro\corref{cor1}}
\ead{pedro.ribeiro@york.ac.uk}
\author{James Baxter}
\ead{james.baxter@york.ac.uk}
\author{Ana Cavalcanti}
\ead{ana.cavalcanti@york.ac.uk}
\address{Department of Computer Science, Univesity of York, UK}





\begin{abstract}
The $tock$-CSP encoding embeds a rich and flexible approach to modelling
discrete timed behaviours in CSP where the event $tock$ is interpreted to mark
the passage of time. The model checker FDR provides tailored support for
$tock$-CSP, including a prioritisation operator that has typically been used
to ensure maximal progress, where time only advances after internal activity
has stabilised. Prioritisation may also be used on its own right as a
modelling construct. Its operational semantics, however, is only congruent
over the most discriminating semantic model of CSP: the finite-linear model.
To enable sound and compositional reasoning in a $tock$-CSP setting, we
calculate a denotational definition for prioritisation. For that we
establish a Galois connection between a specialisation of the finite-linear
model, with $tock$ and $\tick$, that signals termination, as special events,
and $\tick$-$tock$-CSP, a model for $tock$-CSP that captures termination,
deadlines, and is adequate for reasoning about timed refinement.
Our results are mechanised using Isabelle/HOL.
\end{abstract}


\end{frontmatter}



\section{Introduction}\label{sec:introduction}
Process algebras like CSP~\cite{Roscoe2010} enable modelling of reactive
systems via named events that correspond to atomic and instantaneous
interactions of interest. To specify budgets and deadlines, and to reason 
about liveness and safety over time, an explicit notion of time is required.
Roscoe introduced the $tock$-CSP encoding, where an event $tock$ is used
to mark the passage of discrete time. 
Extensive use of $tock$-CSP has been reported~\cite{Evans2000,Kharmeh2011,Isobe2012,Goethel2015,Cavalcanti2019}.

The CSP model-checker FDR~\cite{FDR} has operators tailored for $tock$-CSP.
In addition, it implements a $\mathbf{Pri_{\le}} (P)$ operator that can be used to prioritise
events according to a partial order $\le$. The behaviour is that of $P$, but 
changed so that whenever events $a$ and $b$ are available, then if $b$ is of 
strictly higher priority than $a$, that is, $a < b$, then $a$, and the 
behaviour following $a$, is pruned.
\begin{example}\label{eg:R}%
$R = a \then \mathbf{Skip} \extchoice b \then \mathbf{Skip}$.
Process $R$ offers an external choice ($\extchoice$) between behaving as
a prefixing ($\then$) on $a$ or $b$, followed by immediate termination ($\mathbf{Skip}$).
Prioritising $R$, with $a < b$ yields the process $b \then \mathbf{Skip}$.\qed
\end{example}
Prioritisation can be used in FDR to enforce maximal
progress, that is, that time can only advance after internal activity has
stabilised, by prioritising $\tau$, the internal action, and $\tick$, which
signals termination, over $tock$. $\mathbf{Pri_{\le}}$ also endows CSP with
extra expressivity~\cite{Roscoe2015}, and has been applied also 
in abstraction techniques~\cite{Roscoe2013}, reducing refinement in different CSP semantics
to traces refinement~\cite{Mestel2016}, or as a modelling construct on its own~\cite{Goethel2015,Cavalcanti2019}.

The operator $\mathbf{Pri_{\le}}$ has an intuitive operational 
semantics. However, for $\mathbf{Pri_{\le}}$ to be congruent over denotational models of CSP, namely
the finite-linear ($\mathcal{FL}$~\cite{Roscoe2010}) and refusal testing 
($\mathcal{RT}$~\cite{Phillips1987,Mukarram1993}) models, the partial order needs to be 
constrained~\cite{Roscoe2015}. Thus, FDR actually implements a constrained form of 
$\mathbf{Pri_{\le}}$, where, for example, $\tick$ and $\tau$ are maximal in the order. 
However, this is insufficient for $\mathbf{Pri_{\le}}$
to be congruent over weaker models such as $\tick$-$tock$~\cite{Baxter2019a} and the 
stable-failures ($\mathcal{F}$~\cite{Roscoe2010}). 
\begin{example}\label{example:ST}%
\begin{displaymath}
\begin{aligned}[t]
S &= a \then \mathbf{Skip} \intchoice b \then \mathbf{Skip} \intchoice (\mathbf{Wait}~1~;S)\\
T &= (a \then \mathbf{Skip} \extchoice b \then \mathbf{Skip}) \intchoice (\mathbf{Wait}~1~;T)
\end{aligned}
\end{displaymath}
Process $S$ makes an internal choice ($\intchoice$) between offering events
$a$, or $b$, and terminating, or waiting a time unit ($\mathbf{Wait}~1$) and
then behaving as $S$ again, specified using sequential composition ($~;$). Process
$T$ also makes an internal choice, but the choice between $a$ and $b$ is external.
Although it may seem that $T$ is more deterministic than $S$, deadlocking for a time
unit, before making an internal choice again that may lead to the refusal of $a$ and
$b$, is a possibility for $T$. The $\tick$-$tock$ model does not distinguish $S$
and $T$, just like the $\mathcal{F}$ model does not distinguish
$a \then \mathbf{Skip} \intchoice b \then \mathbf{Skip} \intchoice \mathbf{Stop}$
and $(a \then \mathbf{Skip} \extchoice b \then \mathbf{Skip}) \intchoice \mathbf{Stop}$.

However, prioritising $S$ and $T$, with $a < b$, assuming $\tick$ and $tock$ are 
maximal in the order and that $tock$ is not prioritised over $b$, yields different results.
$T$ becomes equal to a process $priT = b \then \mathbf{Skip} \intchoice (\mathbf{Wait}~1~;priT)$,
whereas the prioritisation of $S$ bears no effect. The incongruence 
arises as a result of $\mathbf{Pri_{\le}}$ being defined over
$\mathcal{FL}$, where it distributes over internal choice. To reason
about priorities in other models we need different definitions for 
$\mathbf{Pri_{\le}}$.\qed
\end{example}
In this paper we calculate a definition of $\mathbf{Pri_{\le}}$ for
$tock$-CSP via a stepwise Galois connection between the $\mathcal{FL}$
model and the $\tick$-$tock$ model~\cite{Baxter2019a}, the only sound model for $tock$-CSP
that can be used to reason about timed refinement, and that captures
deadlines, termination and erroneous Zeno behaviour.
It is a faithful account of the $tock$-CSP dialect and is mechanised in
Isabelle/HOL~\cite{NWP02}. Included in this work is a mechanisation of the $\mathcal{FL}$ model,
that handles termination, via the special event $\tick$, and a specialisation
that includes $tock$.

This paper is structured as follows. In~\cref{sec:tock-CSP} we
describe $\tick$-$tock$ and $\mathcal{FL}$. In~\cref{sec:Formal-FL} 
we formally define $\mathcal{FL}$. In~\cref{sec:Galois-connection} we 
define a Galois connection between a specialisation of $\mathcal{FL}$, 
which includes $tock$ as event, and calculate the induced 
definition of $\mathbf{Pri_{\le}}$. We conclude in~\cref{sec:conclusions}.

\section{Models}
Here we describe the $\tick$-$tock$ and $\mathcal{FL}$ models by summarizing material from~\cite{Baxter2019a} 
and~\cite{Roscoe2010}.

\subsection{$\tick$-$tock$}\label{sec:tock-CSP}

The $\tick$-$tock$ model is defined in terms of a set $\Sigma$ of events, not including
$\tick$ and $tock$. The complete set is defined as $\Sigma^{\tick}_{tock}$.
The semantics of processes is a set of sequences of observations of type
$Obs$, defined below. An observation is either an event in $\Sigma^{\tick}_{tock}$, 
or the refusal of some subset of $\Sigma^{\tick}_{tock}$, so $Obs$ has two 
constructor functions $evt$ and $ref$.
\begin{linkdefinition}{https://github.com/robo-star/tick-tock-CSP/blob/v1/TickTock/TickTock_Core.thy\#L8}
$Obs ::= evt \ldata \Sigma^{\tick}_{tock} \rdata | ref \ldata \power \Sigma^{\tick}_{tock} \rdata$
\end{linkdefinition}
The type of valid traces is defined as $TT$, which is the set of
all sequences $t$ with elements of type $Obs$ that satisfy three conjuncts, defined below.
\begin{linkdefinition}{https://github.com/robo-star/tick-tock-CSP/blob/v1/TickTock/TickTock_Core.thy\#L11}
  $\begin{aligned}[t]
	  &TT == \{ t : \seq Obs | \forall i : \dom t @ \\
                          &\hspace{-6.6em}\begin{aligned}[b]
                          &i < \#t \implies t(i) \neq evt~\tick \land \\
                          &(i < \#t \land t(i) \in \ran ref) \implies t(i+1) = evt~tock \land \\ 
                          &t(i) = evt~tock \implies (i > 1 \land t(i-1) \in \ran ref)
                          \end{aligned}\}
  \end{aligned}$
\end{linkdefinition}
The first conjunct ensures that $\tick$ can only appear as the last event of $t$.
The second conjunct requires that every refusal before the last ($i < \#t$) is followed by a $tock$. Finally, the third conjunct
ensures that every $tock$ is preceded by a refusal. 

The healthiness conditions of $\tick$-$tock$~\cite{Baxter2019a}, whose composition is named
$\mathbf{TT}$, ensure properties of the standard models of CSP in the context
of $TT$ traces. Namely, the empty trace is an observation of every process;
prefix closure and subset closure of refusals; and an event that cannot be
performed is refused. Finally, whenever there is a stable refusal in a trace,
then there is always at least one (other) trace where the refusal includes $\tick$.

\subsection{Finite-linear}
The $\mathcal{FL}$ model is characterised in~\cite{Roscoe2010} by finite sequences $\lseq A_0, e_0,
\ldots, A_i, e_i, A_{i+1} \rseq$ of alternating acceptances $A_i$ and events $e_i$.
An acceptance is either a set, recording the events being offered,
or null ($\bullet$) indicating the impossibility to observe such a
set because of instability. An event $e_i$ necessarily belongs to $A_i$ if $A_i$
is not $\bullet$. Valid sequences end in an acceptance, or
$\bullet$ followed by $\tick$. The healthiness conditions of $\mathcal{FL}$
ensure prefix closure, and require that, whenever an acceptance $A_i$ is
observed, then any events in $A_i$ can also be performed.

The prefix relation for sequences in this model allows $\bullet$ to precede a
stable acceptance set, so that $\lseq \bullet \rseq$ and $\lseq \bullet, e_0, \bullet\rseq$ 
are prefixes of $\lseq A_0, e_0, A_1\rseq$, for example. Crucially, and unlike other CSP models, there is
no upward-closure of acceptance sets. 

\section{Formalising the $\mathcal{FL}$ model}
\label{sec:Formal-FL}

In~\cref{sec:FL:model} we define a recursive data type to
capture $\mathcal{FL}$ traces. In~\cref{sec:FL:healthiness-conditions} we
formalise the healthiness conditions. Finally, in~\cref{sec:FL:Pri} we formally
define $\mathbf{Pri_{\le}}$. While~\citet{Roscoe2010} characterises the
$\mathcal{FL}$ model and studies in depth its relationship to other CSP
models, here we define its data model and healthiness conditions with the
level of formality required to mechanise it (as an Isabelle theory as presented
in~\cite{FLIsabelle}, for example).

\subsection{Model}
\label{sec:FL:model}
An acceptance is either a null acceptance or a set of events. It is
defined by the type $Acc$, which has two constructor functions $\bullet$
and $aset$, respectively, where $\ldata\_\rdata$ is the Z~\cite{WD96}
syntax for constructors.
\begin{linkdefinition}{https://github.com/robo-star/tick-tock-CSP/blob/v1/FL/Finite_Linear_Model.thy\#L17}
$Acc ::= \bullet | aset \ldata \power \Sigma^{\tick}_{tock} \rdata$
\end{linkdefinition}
We also define: $e \in_{\mathcal{FL}} A$ to be $true$ exactly when $A$ is the
result of applying $aset~B$ for some set $B$ and $e \in B$; and a prefix order on
$Acc$, where $\bullet$ is the least element, and $aset~A \le aset~A$, which
can be lifted to define the prefix order on $FL$ traces.

An acceptance followed by an event is a pair of type $Aev$ associating elements of $Acc$ to $\Sigma^{\tick}_{tock}$.
\begin{linkdefinition}{https://github.com/robo-star/tick-tock-CSP/blob/v1/FL/Finite_Linear_Model.thy\#L145}[Acceptance-event]
{\begin{axdef}
	Aev : Acc \times \Sigma^{\tick}_{tock}
	\where
	\forall B : \power \Sigma^{\tick}_{tock}; (aset~B,e) : Aev @ e \in B
\end{axdef}}
\end{linkdefinition}
An acceptance-event pair $(aset~B,e)$ is valid exactly when $e$ is
a member of set $B$. A pair $(\bullet,e)$ imposes no condition on $e$.
Next we define the type of non-empty traces for $\mathcal{FL}$ as $FL$,
a recursive data type with two constructors, $acc$ and $aev$.
\begin{linkdefinition}{https://github.com/robo-star/tick-tock-CSP/blob/v1/FL/Finite_Linear_Model.thy\#L253}
$FL ::= acc \ldata Acc \rdata | aev \ldata Aev \times FL \rdata$
\end{linkdefinition}
The function $acc$ takes a single acceptance while $aev$ takes an 
acceptance-event pair and an $FL$ trace. Processes in $\mathcal{FL}$ are 
defined by a set of $FL$ traces, effectively finite non-empty sequences ending in
an acceptance. Unlike the original presentation of $\mathcal{FL}$, we consider
$\tick$ as a regular event. This simplifies the type definition of $FL$ 
and allows us to give a general account of
$\mathbf{Pri_{\le}}$ where $\tick$ can be prioritised.

To facilitate presentation we abbreviate $acc~A$ as $\lseq A \rseq_{\mathcal{FL}}$,
and $aev~((A,e),\rho)$ as $(A, e)~\#~\rho$. Furthermore a recursive
application of $\#$ a number of times, such as $(A_0,e_0)~\#~(\ldots~\#~A_i)$ 
is abbreviated as $\lseq (A_0,e_0), \ldots, A_i\rseq_{\mathcal{FL}}$. This is a
typical approach to encoding finite lists via recursive data types.

\subsection{Healthiness conditions}
\label{sec:FL:healthiness-conditions}
The healthiness conditions are listed in~\cref{tab:FL:healthiness-conditions}.
The first, $\mathbf{FL0}$, although not listed in~\cite[p.257]{Roscoe2010}, is
required to ensure that every process has some behaviour. Together with
$\mathbf{FL1}$ (adopted from~\cite{Roscoe2010}), which ensures prefix closure,
$\mathbf{FL0}$ ensures that every process has at least the
trace $\lseq\bullet\rseq_{\mathcal{FL}}$. 

The next condition $\mathbf{FL2}$ from \cite{Roscoe2010} is restated using
$\in_{\mathcal{FL}}$ and a concatenation operator $\cat_{\mathcal{FL}}$ that is
closed under $FL$, unlike the general sequence concatenation operator.
$\mathbf{FL2}$ states that whenever a trace $\rho$ concatenated with an
acceptance $A$ is in $P$, then for every event $e$ in $A$ there must be a
trace in $P$ that performs $e$. It is the result of concatenating $\rho$ with the trace
consisting of the acceptance-event pair $(A,e)$ followed by $\bullet$.

To ensure $\tick$ is only possible after $\bullet$ as the very last event of
a trace, $\mathbf{FL3}$ requires valid traces to be in a set $FL_{\tick}$. 
It consists of $FL$ traces where $\tick$ is not offered in any acceptance and, if
$\tick$ appears, it is as the last event followed by $\bullet$.

The $\mathcal{FL}$ model forms a complete lattice under 
the refinement order defined by subset inclusion. The top is the process $\mathbf{div}$
whose only observation is the trace $\lseq\bullet\rseq_{\mathcal{FL}}$, while the bottom is $FL_{\tick}$,
the set of all possible observations: the process $\mathbf{Chaos} (\Sigma)$.

\begin{table}[t!]
	\centering
	\setlength{\abovedisplayskip}{-0.2cm}
	\setlength{\belowdisplayskip}{-0.3cm}
	\setlength{\abovedisplayshortskip}{-0.2cm}
	\setlength{\belowdisplayshortskip}{-0.3cm}
	\setlength{\zedindent}{0cm}
	\setlength{\zedleftsep}{0cm}
	\rowcolors{2}{lightgray}{white}
	\begin{tabular}{p{1.3cm}p{5.6cm}}
		\hline
		Name & Definition \\
		\hline
		\[\href{https://github.com/robo-star/tick-tock-CSP/blob/v1/FL/Finite_Linear_Model.thy\#L694}{\mathbf{FL0}} (P)\] & \[P \neq \emptyset\] \\
		\[\href{https://github.com/robo-star/tick-tock-CSP/blob/v1/FL/Finite_Linear_Model.thy\#L701}{\mathbf{FL1}} (P)\] & \[\rho \leq \sigma \land \sigma \in P \implies \rho \in P\] \\
		\[\href{https://github.com/robo-star/tick-tock-CSP/blob/v1/FL/Finite_Linear_Model.thy\#L718}{\mathbf{FL2}} (P)\] & \[\rho \cat_{\mathcal{FL}} \lseq A\rseq_{\mathcal{FL}} \in P \land e \in_{\mathcal{FL}} A
									\\\implies \rho \cat_{\mathcal{FL}} \lseq (A,e),\bullet\rseq_{\mathcal{FL}} \in P\] \\
		\[\href{https://github.com/robo-star/tick-tock-CSP/blob/v1/FL/Finite_Linear_Tick_Param.thy\#L35}{\mathbf{FL3}} (P)\] & \[P \subseteq FL_{\tick}\] \\
		\hline
	\end{tabular}
	\caption{The healthiness conditions of $\mathcal{FL}$.}
	\label{tab:FL:healthiness-conditions}
\end{table}%

\subsection{Prioritise}
\label{sec:FL:Pri}

The semantics of $\mathbf{Pri_{\le}} (P)$ is defined in~\cite{Roscoe2015} pointwise over
the $FL$ traces of $P$ as follows.
\begin{linkdefinition}{https://github.com/robo-star/tick-tock-CSP/blob/v1/FL/Finite_Linear_Pri.thy\#L30}
\begin{displaymath}
\mathbf{Pri_{\le}} (P) \circdef \{ \rho, \sigma : FL | \rho~\mathbf{pri}_{\le}~\sigma \land \sigma \in P @ \rho \}
\end{displaymath}
\end{linkdefinition}
It is the set of traces $\rho$ related by $\mathbf{pri}_{\le}$ to a trace
$\sigma$ drawn from $P$, that is, a trace $\rho$ is a possible prioritisation
of $\sigma$. The relation $\mathbf{pri}_{\le}$ is a fully specified version of that presented 
in~\cite{Roscoe2015}, generalised to cater for any event, including $\tick$, and 
defined inductively in~\cref{def:pri} as required for mechanical reasoning. 
Observe that $\mathbf{pri}_{p}$ is parametric over an arbitrary partial order $p$ on events.
Before explaining the definition, we illustrate the calculation of traces $\rho$, 
as related by $\mathbf{pri}_{\le}$, from traces $\sigma$ of process $R$ (of \cref{eg:R}),
assuming $a < b$ and that $\tick$ is maximal in the order. The set of traces of $R$ is 
presented below, as well as those of $\mathbf{Pri_{\le}} (R)$, where we omit the
$_{\mathcal{FL}}$ subscripts. We observe that the result is independent of whether
$b$ is maximal in the order.
\begin{example}\label{eg:Pri-R}
{\setlength{\zedindent}{0cm}%
\begin{displaymath}
	traces(R) =\\
 	\left\{\begin{array}{l@{}l@{~}l@{}}
			&\lseq\bullet\rseq, 							&\lseq aset~\{a,b\}\rseq,
		\\	&\lseq(\bullet,a),\bullet\rseq, 				&\lseq(aset~\{a,b\},a),\bullet\rseq,
		\\	&\lseq(\bullet,b),\bullet\rseq,					&\lseq(aset~\{a,b\},b),\bullet\rseq,
		\\	&\lseq(\bullet,a),(\bullet,\tick),\bullet\rseq,	&\lseq(aset~\{a,b\},a),(\bullet,\tick),\bullet\rseq, 
		\\	&\lseq(\bullet,b),(\bullet,\tick),\bullet\rseq, &\lseq(aset~\{a,b\},b),(\bullet,\tick),\bullet\rseq
	\end{array}\right\}
\end{displaymath}}%
We have traces: $\lseq aset~\{a,b\}\rseq$ recording that both events $a$ and $b$ are stably offered, and
traces such as $\lseq(aset~\{a,b\},a),(\bullet,\tick),\bullet\rseq$ recording that after
stably offering both events, and performing event $a$, then termination, encoded
by $\tick$ happens unstably, and similarly for event $b$;
and finally traces that complete the prefix closure as required by $\mathbf{FL1}$.
{\setlength{\zedindent}{0cm}%
\begin{displaymath}
	traces(\mathbf{Pri_{\le}} (R)) =\\
 	\left\{\begin{array}{l@{}l@{~}l@{}}
			&\lseq\bullet\rseq, 							&\lseq aset~\{b\}\rseq,
		\\	&\lseq(\bullet,b),\bullet\rseq,					&\lseq(aset~\{b\},b),\bullet\rseq,
		\\	&\lseq(\bullet,b),(\bullet,\tick),\bullet\rseq, &\lseq(aset~\{b\},b),(\bullet,\tick),\bullet\rseq
	\end{array}\right\}
\end{displaymath}}%
The traces of $\mathbf{Pri_{\le}} (R)$ are those of $b \then \mathbf{Skip}.$\qed
\end{example}
Prioritisation of $R$ with $a < b$, as previously discussed in~\cref{sec:introduction},
leads to pruning the behaviours of the prefixing on $a$. The traces of
$\mathbf{Pri_{\le}} (R)$ are those related to the traces of $R$ by
$\mathbf{pri}_{\le}$.

The trace $\lseq\bullet\rseq$ is related to itself as it is a valid observation
of every process in $\mathcal{FL}$. The trace $\lseq aset~\{b\}\rseq$ is related
to $\lseq aset~\{a,b\}\rseq$, as $\{b\}$ is the set obtained by eliminating 
events in $\{a,b\}$ for which an event of strictly higher priority is also in
the set, namely $a$ is eliminated as $b$ is of higher priority. Similarly, the 
trace $\lseq(aset~\{b\},b),\bullet\rseq$ is related to
$\lseq(aset~\{a,b\},b),\bullet\rseq$, and $\lseq(aset~\{b\},b),(\bullet,\tick),\bullet\rseq$
is related to $\lseq(aset~\{a,b\},b),(\bullet,\tick),\bullet\rseq$ as, in addition,
$\tick$ is maximal. The trace $\lseq(\bullet,b),\bullet\rseq$
is related to $\lseq(aset~\{a,b\},b),\bullet\rseq$ as $\bullet$ is a prefix
of $\{b\}$, and, as discussed, $\{b\}$ is the set obtained from $\{a,b\}$
according to the priority order. Similar observation
applies to trace $\lseq(\bullet,b),(\bullet,\tick),\bullet\rseq$, which is related to
$\lseq(aset~\{a,b\},b),(\bullet,\tick),\bullet\rseq$. Traces
where the event $a$ appears are not related by $\mathbf{pri}_{\le}$ and thus are 
eliminated by $\mathbf{Pri}_{\le} (R)$. Below we define $\mathbf{pri}_{p}$.

\begin{linkdefinition}{https://github.com/robo-star/tick-tock-CSP/blob/v1/FL/Finite_Linear_Pri.thy\#L16}\label{def:pri}
{\setlength{\zedindent}{0cm}
\begin{axdef}
\_~\mathbf{pri}_{\_}~\_ : (\Sigma^{\tick}_{tock} \rel \Sigma^{\tick}_{tock}) \fun (FL \rel FL)
\where
\forall A, Z : Acc~; p : \Sigma^{\tick}_{tock} \rel \Sigma^{\tick}_{tock} @ \\
\lseq A\rseq~\mathbf{pri}_{p}~\lseq Z\rseq \iff A = \mathbf{priacc}_{p} (Z)
\also
((A,e)~\#~\rho)~\mathbf{pri}_{p}~((Z,e)~\#~\sigma) \iff\\
	\quad 
	\left(\begin{array}{l}
		A \leq \mathbf{priacc}_{p} (Z) \land \rho~\mathbf{pri}_{p}~\sigma \land \\
		\lnot max(p,e) \implies e \in_{\mathcal{FL}} \mathbf{priacc}_{p} (Z)
		\end{array}\right)
\end{axdef}}
\end{linkdefinition}
In general, a trace $\lseq A\rseq$ is related to $\lseq Z\rseq$ by $\mathbf{pri}_{p}$, 
whenever $A$ is $\mathbf{priacc}_{p} (Z)$, specified below, that defines
the result of prioritising an acceptance $Z$. In~\cref{eg:Pri-R} above, we have that
$\mathbf{priacc}_{\le} (aset~\{a,b\}) = aset~\{b\}$.
\begin{linkdefinition}{https://github.com/robo-star/tick-tock-CSP/blob/v1/FL/Finite_Linear_Pri.thy\#L9}\label{def:priacc}
 {\setlength{\zedindent}{0cm}
 \begin{axdef}
 \mathbf{priacc}_{\_}~\_ : (\Sigma^{\tick}_{tock} \rel \Sigma^{\tick}_{tock}) \fun Acc \fun Acc
 \where
 \forall p : \Sigma^{\tick}_{tock} \rel \Sigma^{\tick}_{tock}; Z : \power \Sigma^{\tick}_{tock} @ \\
 	\mathbf{priacc}_{p} (\bullet) = \bullet
 	\also
 	\mathbf{priacc}_{p}~(aset~Z) = \\
 	\quad aset~(Z\cap\{ e | \lnot (\exists b @ b \in Z \land e <_{p} b)\})
 \end{axdef}}
\end{linkdefinition}
Formally, $\mathbf{priacc}_{p} (\bullet)$ is $\bullet$, whereas for an acceptance set $Z$,
$\mathbf{priacc}_{p} (aset~Z)$ is the acceptance set of events in $Z$ for which no event $b$
of strictly higher priority, according to the order $p$, exists in $Z$.

A trace $(A,e)~\#~\rho$ is related to $(Z,e)~\#~\sigma$ by $\mathbf{pri}_{p}$
exactly when $A$ is a prefix of the acceptance permitted by prioritisation,
as defined by $\mathbf{priacc}_{p} (Z)$,
$\rho$ is related to $\sigma$ by $\mathbf{pri}_{p}$, and if the event $e$ is not
maximal in the partial order $p$, where $max(p,e) = \lnot \exists x @ e <_{p} x$,
that is, no other event has strictly higher priority than $e$, then $e$ must be in the
acceptance permitted by prioritisation. In other words, events of maximal
priority are never eliminated, whereas those for which an event of higher
priority exists need to be in the resulting acceptance set obtained
via the intersection. In particular, if $Z$ is $\bullet$ and $e$ is not
maximal, then it is not related by $\mathbf{pri}_{p}$. Although our 
definition of $\mathbf{pri}_{p}$ is rather concise, we have \href{https://github.com/robo-star/tick-tock-CSP/blob/v1/FL/Finite_Linear_Priority.thy\#L249}{established}
using our mechanisation that it is equivalent to that of~\citet{Roscoe2015}.

A formal definition for $\mathbf{pri}_{\le}$ suitable for mechanical
reasoning enables key results about
$\mathbf{Pri}_{\le}$ to be established, namely that it is monotonic and closed
under the healthiness conditions. Using our mechanisation we have established the following 
novel results where the subscript $_{\mathcal{FL}}$ indicates that these are operators of 
the $\mathcal{FL}$ model.
\begin{linklemma}{https://github.com/robo-star/tick-tock-CSP/blob/v1/FL/Finite_Linear_Pri_Laws.thy\#L9}\label{lemma:Pri-properties}
\begin{displaymath}
\begin{aligned}[t]
	&\mathbf{Pri}_{\le} \circ \mathbf{Pri}_{\le} (P) = \mathbf{Pri}_{\le} (P)\\
	&\mathbf{Pri}_{\le} (P \intchoice_{\mathcal{FL}} Q) = \mathbf{Pri}_{\le} (P) \intchoice_{\mathcal{FL}} \mathbf{Pri}_{\le} (Q)\\
	&\mathbf{Pri}_{\le} (a \then_{\mathcal{FL}} P) = a \then_{\mathcal{FL}} \mathbf{Pri}_{\le} (P)
\end{aligned}
\end{displaymath}
\end{linklemma}
The first result in~\cref{lemma:Pri-properties} establishes that $\mathbf{Pri}_{\le}$ is idempotent; the second
that it distributes through internal choice; and the third that it distributes through
prefixing. It also distributes through sequential composition, as established next.
\begin{linklemma}{https://github.com/robo-star/tick-tock-CSP/blob/v1/FL/Finite_Linear_Pri_Laws.thy\#L647} Provided $\tick$ has maximal priority,
$\mathbf{Pri}_{\le} (P\circseq Q) = \mathbf{Pri}_{\le} (P)\circseq\mathbf{Pri}_{\le} (Q)$. 
\end{linklemma}
In this case $\tick$ must be maximal in the order, as otherwise the possibility for $P$
to terminate could be eliminated. More interestingly, the prioritisation of an external choice between
prefixings on events $a$ and $b$, where $a < b$, is the prioritisation of the
prefixing on $b$, with any behaviour following on from $a$ pruned, as established by the lemma.
\begin{linklemma}{https://github.com/robo-star/tick-tock-CSP/blob/v1/FL/Finite_Linear_Pri_Laws.thy\#L647} Provided $a < b$,
{\setlength{\zedindent}{0cm}\begin{displaymath}
	\mathbf{Pri}_{\le} (b \then_{\mathcal{FL}} P \extchoice_{\mathcal{FL}} a \then_{\mathcal{FL}} Q) 
	= \mathbf{Pri}_{\le} (b \then_{\mathcal{FL}} P)\\
\end{displaymath}}
\end{linklemma}%
Proof of this, and other key results, is available\footnote{\url{https://github.com/robo-star/tick-tock-CSP}}.

Our formalisation of $\mathcal{FL}$ and $\mathbf{Pri}_{\le}$ are used
next to calculate a definition for $\mathbf{Pri_{\le}}$ in $\tick$-$tock$.

\section{Galois connection}
\label{sec:Galois-connection}
\begin{figure}[h]
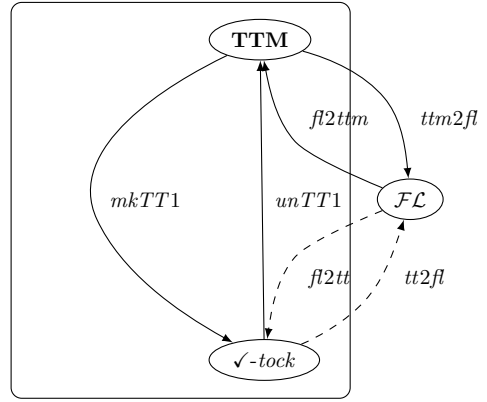

\begin{center}
\begin{dot2tex}[dot,tikz,scale=0.8,mathmode,autosize]
digraph G {
  d2tdocpreamble="\usetikzlibrary{patterns}";

  FL [label="\\mathcal{FL}"];

  FL -> CTM [ label="fl2ttm" ];
  CTM -> FL [ label="ttm2fl" ];
  FL -> CTC [ label="fl2tt", style="dashed"];
  CTC -> FL [ label="tt2fl", style="dashed"];

  subgraph clusterTT {

    graph [style="rounded corners"];
  
    CTM [ label="\\mathbf{TTM}" ];
    CTC [ label="\\tick\text{-}tock" ];

    CTM -> CTC [ label="mkTT1" ];
    CTC -> CTM [ label="unTT1" ];
  }
}
\end{dot2tex}
\end{center}\vspace{-1em}%
\caption{\label{fig:fl-tick-tock}Galois connection between $\tick$-$tock$ and $\mathcal{FL}$ models.}%
\end{figure}

The key to defining a Galois connection between $\mathcal{FL}$ and
$\tick$-$tock$ is relating $FL$ traces, which record acceptances before every
event, and $TT$ traces, which instead record subset-closed refusals,
and only before $tock$ events or at the end of a trace, as previously discussed
in~\cref{sec:tock-CSP}. We define the Galois connection stepwise to simplify proofs. In~\cref{sec:galois:fl-to-tt} 
we first consider a Galois connection between $\mathcal{FL}$ and a variant of
$\tick$-$tock$ where refusals sets are not subset closed; instead they are maximal: they record
exactly what is refused. In the following~\cref{sec:galois:tt-to-fl} we define a Galois
connection with full $\tick$-$tock$ by completing the subset closure of refusals. 
\cref{fig:fl-tick-tock}, which is explained as we discuss each step, provides a
depiction of the connections. Finally, in \cref{sec:galois:prioritise}, we
present the result of calculating the induced definition of $\mathbf{Pri}_{\le}$ for
$\tick$-$tock$.

\subsection{From $\mathcal{FL}$-$tock$ to maximal $\tick$-$tock$}
\label{sec:galois:fl-to-tt}
The pair of functions $fl2ttm$ and $ttm2fl$, mapping between $\mathcal{FL}$-$tock$ and the
maximal variant of $\tick$-$tock$, labelled as $\mathbf{TTM}$ in~\cref{fig:fl-tick-tock}, is defined below.
\begin{linkdefinition}{https://github.com/robo-star/tick-tock-CSP/blob/v1/TickTock-FL/TickTock_FL.thy\#L11}%
{\setlength{\zedindent}{0cm}%
\begin{displaymath}%
\begin{aligned}
	\href{https://github.com/robo-star/tick-tock-CSP/blob/master/TickTock-FL/TickTock_Max_FL.thy\#L170}{fl2ttm}(P) &\defs \{ \rho : FL | \rho \in P @ fl2ttobs(\rho) \}\\
	\href{https://github.com/robo-star/tick-tock-CSP/blob/master/TickTock-FL/TickTock_Max_FL.thy\#L298}{ttm2fl}(P) &\defs \bigcup \left\{ Q |
			  fl2ttm(Q) \subseteq P
			  \right\}
\end{aligned}\vspace{-0.1em}%
\end{displaymath}}%
\end{linkdefinition}
The function $fl2ttm$ is defined as the set of traces obtained by applying
$fl2ttobs$, a total non-injective function from $FL$ to $TT$ traces,
defined next, to every trace $\rho$ in $P$. The inverse
mapping $ttm2fl$ is uniquely defined in terms of $fl2ttm$, and 
is the distributed union over the set of $\mathbf{FL}$ processes 
that can be mapped via $fl2ttm$ to be a subset of $P$. This construction is standard for Galois
connections where one adjoint uniquely defines the other. 

The function $fl2ttobs$ is defined as follows.
\begin{linkdefinition}{https://github.com/robo-star/tick-tock-CSP/blob/v1/TickTock-FL/TickTock_Max_FL.thy\#L26}%
{\setlength{\zedindent}{0cm}%
\begin{axdef}%
fl2ttobs : FL \fun TT
\where
\forall a : Acc~; e : \Sigma^{\tick}; A : \power \Sigma^{\tick}_{tock} @ \\
	fl2ttobs (\lseq\bullet\rseq_{\mathcal{FL}}) = \lseq\rseq \\
	fl2ttobs (\lseq aset~A\rseq_{\mathcal{FL}}) = \lseq ref~\{ z | z \notin A\}\rseq\\
	fl2ttobs ((a,evt~e) ~\#~\rho) = \lseq evt~e \rseq\cat fl2ttobs(\rho) \\
	fl2ttobs ((\bullet,evt~tock) ~\#~\rho) = \lseq\rseq\\
	fl2ttobs ((aset~A,evt~tock) ~\#~\rho) =\\
		\quad\lseq ref~\{ z | z \notin A\}, evt~tock\rseq\cat fl2ttobs(\rho)
\end{axdef}}%
\end{linkdefinition}
It maps: the null trace $\lseq\bullet\rseq_{\mathcal{FL}}$ to the empty trace $\lseq\rseq$,
since $\tick$-$tock$ does not record instability;
and every trace whose only element is an acceptance set $aset~A$ to a singleton trace
consisting of a refusal obtained as the set complement of $A$. The mapping of an
acceptance-event pair followed by a trace $\rho$ is split into three cases:
(1) acceptances $a$ preceding regular events $e$, other than $tock$, are mapped to a trace
$\lseq evt~e\rseq$ whose only element is the event $e$, concatenated with the result
of applying $fl2ttobs$ to $\rho$, since $\tick$-$tock$ does not record refusals before
such events; (2) for the same reason, null acceptances preceding $tock$ are mapped to the
empty trace; (3) acceptance sets preceding $tock$ are mapped to a trace consisting
of a refusal set, obtained as the complement of $A$, followed by the event $tock$, 
concatenated with the application of $fl2ttobs$ to $\rho$.

The functional application of $fl2ttm$ to a healthy process $P$ satisfies nearly all 
the healthiness conditions of $\tick$-$tock$, as established by~\cref{lemma:fl2ttm:closure}.
It also satisfies extra healthiness conditions, defined below. They are relevant to obtain
the final Galois connection in the next section.
\begin{linklemma}{https://github.com/robo-star/tick-tock-CSP/blob/v1/TickTock-FL/TickTock_Max_FL.thy\#L2331}\label{lemma:fl2ttm:closure}
Provided $P$ satisfies $\mathbf{FL0}$-$\mathbf{3}$, $fl2ttm(P)$ satisfies $\mathbf{TT0}$, $\mathbf{TT1w}$,
$\mathbf{TT2}$-$\mathbf{4}$ and $\mathbf{TTM1}$-$\mathbf{3}$.
\end{linklemma}
Namely, $fl2ttm(P)$ satisfies $\mathbf{TT0}$ and $\mathbf{TT2}$-$\mathbf{4}$, but not $\mathbf{TT1}$~\cite{Baxter2019a}, 
which requires subset closure of refusals. This is expected, as $fl2ttobs$ is a function.
(The subset closure of refusals is completed by the connection in~\cref{sec:galois:tt-to-fl}.) 
Instead, we have prefix-closure of sequences, ensured by a healthiness
condition we call $\mathbf{TT1w}$, a weaker form of $\mathbf{TT1}$ that does not
enforce subset closure of refusals. 

In addition, refusals are maximal, and so $fl2ttm(P)$ satisfies the additional
conditions $\mathbf{TTM1}$-$\mathbf{3}$ listed in~\cref{tab:max-tick-tock:healthiness-conditions}.
$\mathbf{TTM1}$ requires that every event $e$ that is not refused
can be performed. $\mathbf{TTM2}$ is similar, but requires $tock$ to happen
after the refusal. $\mathbf{TTM3}$ requires that $P$ is a subset of
$TT_{\tick}$, the set of all traces where every refusal contains $\tick$.
The three conditions, together named $\mathbf{TTM}$ (see~\cref{fig:fl-tick-tock}), characterise the set of
$\tick$-$tock$ processes whose refusal sets contain exactly the events refused.

Since $fl2ttm$ is monotonic with respect to the refinement order, and $ttm2fl$ 
is closed under the healthiness conditions of $\mathbf{FL}$, we have a Galois 
connection between $\mathcal{FL}$-$tock$ and maximal $\tick$-$tock$. 

Next we focus on the subset closure of refusals to complete the Galois connection.

\begin{table}[t!]
	\centering
	\setlength{\abovedisplayskip}{-0.2cm}
	\setlength{\belowdisplayskip}{-0.3cm}
	\setlength{\abovedisplayshortskip}{-0.2cm}
	\setlength{\belowdisplayshortskip}{-0.3cm}
	\setlength{\zedindent}{0cm}
	\setlength{\zedleftsep}{0cm}
	\rowcolors{2}{lightgray}{white}
	\begin{tabular}{p{1.5cm}p{5.6cm}}
		\hline
		Name & Definition \\
		\hline
		\[\href{https://github.com/robo-star/tick-tock-CSP/blob/v1/TickTock-FL/TickTock_Max.thy\#L11}{\mathbf{TTM1}} (P)\] & \[\rho \cat \lseq ref~X\rseq \in P \land e \notin X \land e \neq tock \\
									\implies \rho\cat\lseq evt~e\rseq \in P\] \\
		\[\href{https://github.com/robo-star/tick-tock-CSP/blob/v1/TickTock-FL/TickTock_Max.thy\#L32}{\mathbf{TTM2}} (P)\] & \[\rho \cat \lseq ref~X\rseq \in P \land tock \notin X \\
									\implies \rho\cat\lseq ref~X,evt~tock\rseq \in P\] \\
		\[\href{https://github.com/robo-star/tick-tock-CSP/blob/v1/TickTock-FL/TickTock_Max.thy\#L122}{\mathbf{TTM3}} (P)\] & \[P \subseteq TT_{\tick}\]\\
		\hline
	\end{tabular}
	\caption{The healthiness conditions of maximal $\tick$-$tock$.}
	\label{tab:max-tick-tock:healthiness-conditions}
\end{table}%

\subsection{Subset-closure of refusals leading to full $\tick$-$tock$}
\label{sec:galois:tt-to-fl}

Completing the subset closure of refusal sets in $\tick$-$tock$ is achieved
by the function $mkTT1$.
%
\begin{linkdefinition}{https://github.com/robo-star/tick-tock-CSP/blob/v1/TickTock-FL/TickTock_FL.thy\#L11}%
{\setlength{\zedindent}{0cm}
\begin{displaymath}
\begin{aligned}
	\href{https://github.com/robo-star/tick-tock-CSP/blob/v1/TickTock/TickTock_Core.thy\#L572}{mkTT1}(P) &\defs P \cup \{ \rho, \sigma : TT | \rho \lesssim \sigma \land \sigma \in P @ \rho \}\\
	\href{https://github.com/robo-star/tick-tock-CSP/blob/v1/TickTock-FL/TickTock_Max_TT1.thy\#L149}{unTT1}(P) &\defs \bigcup \left\{ Q |
										   mkTT1(Q) \subseteq P
										\right\}
\end{aligned}
\end{displaymath}}%
\end{linkdefinition}
With $mkTT1(P)$, we have the union of $P$ with a set of traces $\rho$ related by $\lesssim$,
the prefix relation on $\tick$-$tock$ traces defined in~\cite{Baxter2019a}, to $\sigma$ drawn from
$P$. The other adjoint is $unTT1$ and is uniquely defined in terms of $mkTT1$. It is the
distributed union over processes $Q$ (where $Q$ satisfies $\mathbf{TTM}$ and $\mathbf{TT1w}$)
and whose mapping via $mkTT1(Q)$ is a subset of $P$. In other words, $unTT1$ undoes the subset
closure of refusals resulting from $mkTT1$, yielding a process whose refusals are maximal,
as defined by $\mathbf{TTM}$, and whose traces are prefix-closed under $\mathbf{TT1w}$.

Importantly, as stated below, $mkTT1$ preserves healthiness of processes that satisfy $\mathbf{TT0}$ and
$\mathbf{TT2}$-$\mathbf{4}$, and is $\mathbf{TT1}$-healthy as $mkTT1(P) = P$ if, and only if,
$\mathbf{TT1} (P)$.
%
\begin{linklemma}{https://github.com/robo-star/tick-tock-CSP/blob/v1/TickTock-FL/TickTock_Max_TT1.thy\#L293}[Closure]
Provided $P$ satisfies $\mathbf{TT0}$, $\mathbf{TT2}$-$\mathbf{4}$, $\mathbf{TT1w}$
and $\mathbf{TTM1}$-$\mathbf{3}$, then $mkTT1(P)$ is a $\tick$-$tock$ process, that is, it
satisfies $\mathbf{TT}$.
\end{linklemma}
%
Thus, we have a Galois connection between maximal $\tick$-$tock$, labelled as $\mathbf{TTM}$ in~\cref{fig:fl-tick-tock},
and full $\tick$-$tock$. Next, we compose both connections to obtain a Galois 
connection between $\mathcal{FL}$ and $\tick$-$tock$, depicted in~\cref{fig:fl-tick-tock}
by the dashed arrows.
\begin{linkdefinition}{https://github.com/robo-star/tick-tock-CSP/blob/v1/TickTock-FL/TickTock_FL.thy\#L12}
$\begin{aligned}[t]
	\href{https://github.com/robo-star/tick-tock-CSP/blob/v1/TickTock-FL/TickTock_FL.thy\#L14}{fl2tt}(P) \defs mkTT1\circ fl2ttm(P)\\
	\href{https://github.com/robo-star/tick-tock-CSP/blob/v1/TickTock-FL/TickTock_FL.thy\#L16}{tt2fl}(P) \defs ttm2fl\circ unTT1(P)\\
\end{aligned}$
\end{linkdefinition}
The function $fl2tt(P)$ maps $\mathcal{FL}$ processes to $\tick$-$tock$. In the opposite 
direction we have $tt2fl$.

\begin{figure*}[!b]%
\stepcounter{definition}%
\begin{linkdefinition}{https://github.com/robo-star/tick-tock-CSP/blob/v1/TickTock-FL/TickTock_FL_Pri.thy\#L183}%
{\setlength{\zedindent}{0cm}
\begin{axdef}
\_\mathrel{\mathbf{pri_{TT}}_{[\_,\_,\_]}}\_ : (\Sigma^{\tick}_{tock} \rel \Sigma^{\tick}_{tock}) \times \seq Obs \times \power TT \fun (\seq Obs \rel \seq Obs)
\where
\forall p : \Sigma^{\tick}_{tock} \rel \Sigma^{\tick}_{tock}; 
		e : \Sigma^{\tick}_{tock}; 
		S, R : \power \Sigma^{\tick}_{tock}; 
		\rho_0,\rho_1,\sigma \seq Obs;
		Q : \power TT @ \\
\encircle{\scriptsize 1}\; \lseq\rseq\mathrel{\mathbf{pri_{TT}}_{[p,\sigma,Q]}}\lseq\rseq
\also\encircle{\scriptsize 2}\;
\lseq ref~R\rseq\mathrel{\mathbf{pri_{TT}}_{[p,\sigma,Q]}}\lseq ref~S\rseq 
	\iff R \subseteq \mathbf{priref}(p,\sigma,Q,S)
\also\encircle{\scriptsize 3}\;
(\lseq ref~R, evt~tock\rseq \cat \rho_0)\mathrel{\mathbf{pri_{TT}}_{[p,\sigma,Q]}}(\lseq ref~S, evt~tock\rseq \cat \rho_1) \\
	\iff R \subseteq \mathbf{priref}(p,\sigma,Q,S) \land tock \notin \mathbf{priref}(p,\sigma,Q,S) \land
				\rho_0\mathrel{\mathbf{pri_{TT}}_{[p,\sigma\cat\lseq ref~S, evt~tock\rseq,Q]}}\rho_1
\also\encircle{\scriptsize 4}\;
(\lseq evt~e\rseq\cat\rho_0)\mathrel{\mathbf{pri_{TT}}_{[p,\sigma,Q]}}(\lseq evt~e\rseq\cat\rho_1)\\
	\iff 
		\rho_0\mathrel{\mathbf{pri_{TT}}_{[p,\sigma\cat\lseq evt~e\rseq,Q]}}\rho_1 \land
		\left(\lnot max(p,e) \implies
		(e\neq\tick \land \exists Z @ \sigma\cat\lseq ref~Z\rseq \in Q\land e\notin \mathbf{priref}(p,\sigma,Q,Z))
		\right)
\end{axdef}}\label{def:priTT}\vspace{-3em}
\end{linkdefinition}
\end{figure*}

\subsection{Prioritise for $\tick$-$tock$}
\label{sec:galois:prioritise}

Similar to the definition of $\mathbf{Pri_{\le}}$, which is pointwise over $FL$ traces using a relation $\mathbf{pri}$,
we have a similar characterisation for $\mathbf{Pri_{TT}}_{\le}$ as defined next.
\addtocounter{definition}{-2}
\begin{linkdefinition}{https://github.com/robo-star/tick-tock-CSP/blob/v1/TickTock-FL/TickTock_FL_Pri.thy\#L208}\label{def:PriTT}
{\setlength{\zedindent}{0cm}
\begin{displaymath}
	\mathbf{Pri_{TT}}_{\le} (P) = \{
			 \rho, \sigma |
			 	 \rho\mathrel{\mathbf{pri_{TT}}_{[\le,\lseq\rseq,P]}}\sigma
			 	 \land \sigma \in P @ \rho 
			 		\}
\end{displaymath}}
\end{linkdefinition}
\addtocounter{definition}{+1}
The traces of $\mathbf{Pri_{TT}}_{\le}$ is the set of $TT$ traces $\rho$ related to a trace $\sigma$,
from $P$, by a parametric relation $\mathbf{pri_{TT}}_{[\le,\lseq\rseq,P]}$,
specified in \cref{def:priTT}, that takes three parameters: the partial
order $\le$ over events in $\Sigma^{\tick}_{tock}$; the empty
sequence $\lseq\rseq$, and $P$. 

For a trace $\sigma$, prioritisation needs to take into account, at each
point in $\sigma$ where an event is present, the refusal sets that can be
observed just before it, as well as whether other events can be performed.
So the second parameter of $\mathbf{pri_{TT}}$ 
is the trace considered so far in the process specified as the third
parameter. In the initial case,
when considering $\sigma$, the trace considered so far is the empty sequence.
Before explaining the formal definition, we illustrate the calculation of
$\mathbf{Pri_{TT}}_{\le} (T)$ (from \cref{example:ST}), where $a < b$, $\tick$ and $tock$
have maximal priority, and $\lnot b < tock$.
\begin{example}
We first enumerate traces of $T$, with $\Sigma = \{a,b\}$.
{\setlength{\zedindent}{0cm}%
\begin{displaymath}
	traces(T) =\\
 	\left\{\begin{array}{l@{}l@{~}l@{}}
			&\lseq\rseq, 								&\lseq ref~\{\tick\}\rseq,
		\\	&\lseq evt~a\rseq, 							&\lseq evt~b\rseq,
		\\	&\lseq evt~a, evt~\tick\rseq, 				&\lseq evt~b, evt~\tick\rseq,
		\\  &\lseq ref~\{a,b,\tick\}\rseq, 				&\lseq ref~\{a,b,\tick\},evt~tock\rseq,\ldots
	\end{array}\right\}
\end{displaymath}}%
In addition to the empty trace $\lseq\rseq$, we focus on traces containing maximal
refusal sets. The trace $\lseq ref~\{\tick\}\rseq$ stems from the timed external choice, in that
every event not in the refusal set, such as $a$, $b$ and $tock$ is possible.
The traces $\lseq evt~a\rseq$ and $\lseq evt~b\rseq$ capture the possibility to
perform either event, and the traces $\lseq evt~a, evt~\tick\rseq$ and $\lseq evt~b, evt~\tick\rseq$
capture the termination that follows after each event.
The possibility to keep delaying the choice is captured by traces starting with
$\lseq ref~\{a,b,\tick\},evt~tock\rseq$, where every event other than $tock$ is observed 
to be refused, followed by a $tock$ and possibly one of the illustrated
traces above. This refusal set is what distinguishes process $T$ from $R$ in $\tick$-$tock$, as in the case of 
$R$ it is not possible to refuse both events $a$ and $b$ at any time. ($R$ from
\cref{eg:R} refines $T$, by resolving the internal choice in $T$.)\qed
\end{example}
The traces of $\mathbf{Pri_{TT}}_{\le} (T)$ are those related to the traces of $T$
by $\mathbf{pri_{TT}}_{[\le,\lseq\rseq,T]}$, as follows.

The trace $\lseq\rseq$ is related to itself as it is a valid trace of every
process (see~\encircle{\scriptsize 1} in~\cref{def:priTT}). The trace $\lseq ref~\{\tick\}\rseq$ is related to itself, and every
trace whose only refusal is a subset of $\{\tick\}$ is also related to it
because of subset closure (see~\encircle{\scriptsize 2}).
Similarly, $\lseq ref~\{a,b,\tick\}\rseq$ is related
to itself, and to traces whose only refusal is a subset of $\{a,b,\tick\}$.
The trace $\lseq evt~a\rseq$ is equally related to itself
(see~\encircle{\scriptsize 4}), as
although $a$ is not maximal, the presence of the trace $\lseq
ref~\{a,b,\tick\}\rseq$ indicates that $b$ could be refused, in which case $a$
is possible on its own, and thus prioritisation has no effect. Similarly $\lseq
evt~b\rseq$ is related to itself as $b$ is prioritised over $a$, $\tick$
can only happen unstably, and $tock$ is assumed not to be prioritised over
$b$. The traces $\lseq evt~b, evt~\tick\rseq$ and $\lseq evt~a,
evt~\tick\rseq$ are similarly related to themselves, 
given that $\tick$ has maximal priority. Finally, because $tock$ is maximal, $\lseq
ref~\{a,b,\tick\},evt~tock\rseq$ is related to itself (see~\encircle{\scriptsize 3}).

Because $b$ may be refused by $T$ if the choice is delayed as a result of the
internal choice being resolved to $\mathbf{Wait}~1\circseq T$, no traces of $T$ are 
pruned by the application of $\mathbf{Pri_{TT}}_{\le} (T)$ (as they are by the application of
$\mathbf{Pri}_{\le} (T)$ as explained in~\cref{example:ST}). 

Central to the definition of $\mathbf{pri_{TT}}_{[p,\sigma,Q]}$ is a function
$\mathbf{priref} (p,\sigma,Q,S)$, which considers the effect of prioritisation
over a refusal set $S$ (see cases~\encircle{\scriptsize 2} to \encircle{\scriptsize 4} of~\cref{def:priTT}).
A trace $\lseq ref~R\rseq$ is related to $\lseq ref~S\rseq$ by $\mathbf{pri_{TT}}_{[p,\sigma,Q]}$
whenever $R$ is a subset of $\mathbf{priref} (p,\sigma,Q,S)$, which is specified below.
It defines a refusal set containing all the events of $S$ as well as those 
disallowed by prioritisation, according to the order $p$, given that the trace
preceding $\lseq ref~S\rseq$ in $Q$ is $\sigma$. For example, the result of
applying $\mathbf{priref} (p,\lseq\rseq,T,\emptyset)$ is $\{a\}$, assuming
$a <_{p} b$, as event $b$ can be performed in $T$, but is of higher
priority than $a$ and $b$ is not in the empty set. Initially, in the context
of $\mathbf{Pri_{TT}}$, $\sigma$ is the empty trace.
\begin{linkdefinition}{https://github.com/robo-star/tick-tock-CSP/blob/v1/TickTock-FL/TickTock_FL_Pri.thy\#L178}\label{def:priref}
{\setlength{\zedindent}{0cm}
\begin{displaymath}
\mathbf{priref} (p,\sigma,Q,S) = \\
	\quad\begin{aligned}
	S 	&\cup \{ e | \sigma\cat\lseq ref~S, evt~tock\rseq \in Q \land e <_{p} tock\}\\
		&\cup \left\{ e \left| \begin{array}{l}
					\exists b @ b \notin S \land \sigma\cat\lseq evt~b\rseq\in Q\\
					\land b\neq tock\land b\neq \tick \land e <_{p} b
					\end{array}\right.\right\}
	\end{aligned}
\end{displaymath}}
\end{linkdefinition}
The function $\mathbf{priref}$ is defined as the union of $S$ and two sets:
the set of events $e$ of strictly lower priority than $tock$, according to the
order $p$, if the trace $\sigma$ concatenated with the trace where the refusal
$S$ followed by $tock$ is in $Q$; and the set of events $e$ such that there is
some event $b$ of strictly higher priority than $e$, which is not $tock$ nor
$\tick$ and is not in $S$, but that can be performed in $Q$ after trace
$\sigma$ as $\sigma\cat\lseq evt~b\rseq$. Thus, to ascertain whether an
event $e$ can be refused because of 
prioritisation, it is necessary to determine whether an event of higher 
priority, other than $\tick$, that is not refused in $S$ can be performed after trace $\sigma$. The reason
$\tick$ is excluded from the comparison is because it is an unstable event, and
so it is never in a genuine offer to the environment. To illustrate this
point, we consider the process in the following example.
\begin{example}
$K = \mathbf{Skip} \intchoice f \then \mathbf{Skip}$.
In addition to $e$ being offered stably, we have the possibility that process
$K$ terminates. Assuming $f <_{p}\tick$, applying $\mathbf{priref} (p,\lseq\rseq,K,\emptyset)$ would result
in the set $\{f\}$ if $f$ could be compared with $\tick$ 
in $\mathbf{priref}$.
Because $\tick$ is never a genuine option for the environment, however, it is 
incorrect to compare it with events of lower priority.  Moreover, as illustrated 
in this example, the set of traces induced by such a definition of 
$priref(p,\lseq\rseq,K,\emptyset)$ would be unhealthy, as no refusal set
${f,\tick}$ could be obtained by applying $\mathbf{priref}$ to an initial refusal of $K$ (see~\cref{sec:tock-CSP}). 
\end{example}

Similarly, a trace $\lseq ref~R, evt~tock\rseq \cat \rho_0$ is related to
$\lseq ref~S, evt~tock\rseq \cat \rho_1$ by $\mathbf{pri_{TT}}_{[p,\sigma,Q]}$ (see \encircle{\scriptsize 3})
whenever $R$ is a subset of $\mathbf{priref} (p,\sigma,Q,S)$, 
and, in addition, $tock$ is not refused as the result of applying $\mathbf{priref}$, and the traces
$\rho_0$ and $\rho_1$ are related by $\mathbf{pri_{TT}}_{[p,\sigma\cat\lseq ref~S, evt~tock\rseq,Q]}$ where the second
parameter is $\sigma$ concatenated with $\sigma\cat\lseq ref~S, evt~tock\rseq$. For
example, assuming $tock <_{p} a$ then the result of 
$\mathbf{priref} (p,\lseq\rseq,T,\{a,b,\tick\})$ would be
$\{a,b,\tick,tock\}$ because the trace $\lseq evt~a\rseq$
is in $T$, and thus the trace $\lseq ref~\{a,b,\tick\},evt~tock\rseq$ would be pruned by $\mathbf{pri_{TT}}$.

The final case~\encircle{\scriptsize 4} in~\cref{def:priTT} specifies that a trace $\lseq evt~e\rseq\cat\rho_0$
is related to $\lseq evt~e\rseq\cat\rho_1$ by $\mathbf{pri_{TT}}_{[p,\sigma,Q]}$, whenever
$\rho_0$ is related to $\rho_1$ by $\mathbf{pri_{TT}}_{[p,\sigma\cat\lseq evt~e\rseq,Q]}$ and,
if $e$ is not maximal in the priority order $p$, then it is not $\tick$ and
there is some refusal $Z$ observable after $\sigma$ in $Q$, such that application
of $\mathbf{priref} (p,\sigma,Q,Z)$ does not lead to $e$ being refused. For example,
trace $\lseq evt~a\rseq$ from process $T$ is related to itself when considering
$a <_{p} b$, as although $a$ is not maximal there is a trace $\lseq ref~\{b\} \rseq$
in $T$, by subset closure, where the application of $\mathbf{priref} (p,\lseq\rseq,T,\{b\})$ yields $\{b\}$.
In words, if $e$ is maximal then the trace is related, and in particular, if $\tick$
appears in a trace it must be maximal. Otherwise, if $e$ 
is neither maximal nor $\tick$, then it must be the case that application of
$\mathbf{priref}$ to $Q$ indicates that $e$ can be chosen; not refused.

Thus, events that are not maximal are not discarded by prioritisation if, at
the same time, events of higher priority can be refused. We consider, an 
alternative presentation of $T$, by distributing the internal choice
over the external choice.%
\begin{displaymath}%
T = \left(\begin{aligned}
		&(a \then \mathbf{Skip} \intchoice (\mathbf{Wait}~1\circseq T)) 
		\\ \extchoice~ 
		&(b \then \mathbf{Skip} \intchoice (\mathbf{Wait}~1\circseq T))
	\end{aligned}\right)
\end{displaymath}%
This clearly shows the possibility for $a$ to be stably offered for a time
unit, if the first internal choice is decided in favour of $a$, while the
second internal choice is made in favour of a delay. Thus, before the first
time unit has elapsed, there is the possibility that no $b$ is on offer, and
thus prioritisation is ineffective at that point. Another presentation of
$T$ is exactly $S$, which, as discussed in~\cref{sec:introduction}, has the same
semantics as $T$ in $\tick$-$tock$, but not in $\mathcal{FL}$. This is inherently 
related to algebraic properties enjoyed by $\tick$-$tock$ processes, which are 
consistent with the $\mathcal{F}$ model within a time unit. 

Our main result is that the $\mathbf{Pri_{TT}}_{\le}$ operator of $\tick$-$tock$ 
is exactly that induced by the Galois connection with $\mathcal{FL}$ as established
next.
\begin{linkthm}{https://github.com/robo-star/tick-tock-CSP/blob/v1/TickTock-FL/TickTock_FL.thy\#L197}\label{lemma:PriTT}%
Provided $P$ is $\mathbf{TT}$-healthy,%
\begin{displaymath}
\mathbf{Pri_{TT}}_{\le} (P) \defs fl2tt\circ\mathbf{Pri}_{\le} \circ tt2fl(P)
\end{displaymath}%
\end{linkthm}%
For a $\tick$-$tock$ process $P$, $\mathbf{Pri_{TT}}_{\le} (P)$ can be obtained by mapping $P$ into $\mathcal{FL}$, via
$tt2fl$, then applying the prioritisation operator of $\mathcal{FL}$, and finally mapping the result
into $\tick$-$tock$ via $fl2tt$. 

\section{Conclusions}
\label{sec:conclusions}

To endow $\tick$-$tock$ with a prioritise operator consistent with
$\mathbf{Pri_{\le}}$ of $\mathcal{FL}$, we have established a Galois
connection between $\mathcal{FL}$ and $\tick$-$tock$, and calculated an
induced definition. With that, we have a definition that is by construction 
monotonic and healthy. It is strictly not pointwise because prioritisation of 
events in a trace needs to take into account other refusal sets
reachable at the same point. 

The formalisation of $\mathcal{FL}$, although presented here 
in terms of a set $\Sigma^{\tick}_{tock}$, is fully parametric in our
mechanisation in Isabelle/HOL, and thus depends only on $\Sigma$. The 
mechanisation of the $\mathcal{FL}$ model is
layered. We have a faithful mechanisation of $\mathcal{FL}$, built stepwise
to consider termination, and discrete time, by including $\tick$ and $tock$
incrementally.

The algebraic laws from~\cref{sec:Formal-FL} for
$\mathbf{Pri_{\le}}$ are novel, and more importantly are applicable to
$\mathcal{RT}$ as a consequence of~\citet{Roscoe2010}'s results on the
hierarchy of CSP semantic models. Congruence of the operational semantics of
$\mathbf{Pri_{\le}}$ over that model, however, requires further restrictions
to the order~\cite{Roscoe2015}, including that $\tick$ is maximal, enforced, 
for example, when model-checking with FDR.

Effectively, $\mathbf{Pri_{TT}}_{\le}$ can look into
nondeterministic choices unlike other CSP operators.
This presents challanges for its implementation in FDR. It is known
that $\mathbf{Pri_{\le}}$ cannot be specified via combinators, as used in
FDR, and instead an extended theory of combinators is
required~\cite{Roscoe2015}. Thus, $\mathbf{Pri_{\le}}$
is actually implemented in FDR as a function that operates over an LTS, rather than via combinator semantics.
A practical strategy for implementing $\mathbf{Pri_{TT}}_{\le}$ in FDR,
for example, may require some form of
bisimulation, or search, to examine refusals reachable by sequences
of $\tau$ transitions from each state being prioritised in the LTS.

A denotational definition for $\mathbf{Pri_{TT}}$
enables algebraic properties to be explored. It is in our plans to propose laws
in support of a refinement approach to verification of robotic simulations~\cite{Cavalcanti2019}, where prioritisation
is used as part of capturing the assumptions routinely made by roboticists
in simulations.

Finally, the semantic model $\tick$-$tock$ is somewhere between $\mathcal{F}$
and $\mathcal{RT}$ in terms of expressivity, as indicated by the way refusals are recorded.
Our results are likely to be enlightening in the calculation of a counterpart 
of $\mathbf{Pri_{\le}}$ for $\mathcal{F}$ as well.

\section*{Acknowledgments}
This work is funded by the EPSRC grants EP/M025756/1 and EP/R025479/1, and by
the Royal Academy of Engineering.  No new primary data was created as part of
the study reported here.


\bibliography{reading}

\end{document}